\documentclass{article}
\usepackage{spconf,amsmath,amssymb,graphicx}
\usepackage{multirow}
\usepackage[table,xcdraw]{xcolor}
\usepackage{lipsum}
\usepackage[export]{adjustbox}
\usepackage{mathtools}
\usepackage[mathscr]{euscript}
\usepackage[utf8]{inputenc}
\usepackage{hyperref}
\usepackage{amsmath}

\usepackage{algorithm}
\usepackage{algcompatible}
\usepackage{float}

\DeclarePairedDelimiter{\ceil}{\lceil}{\rceil}
\DeclareMathOperator{\argmin}{\arg\!\min}

\algnewcommand\algorithmicreturn{\textbf{return}}
\algnewcommand\RETURN{\State \algorithmicreturn}
\title{MULTISCALE NAKAGAMI PARAMETRIC IMAGING FOR \\IMPROVED LIVER TUMOR LOCALIZATION}
\name{Omar S. Al-Kadi}%\thanks{Thanks to XYZ agency for funding.}}
\address{King Abdullah II School for IT\\ University of Jordan\\ Amman 11942, Jordan}
\begin{document}
%\ninept
%
\maketitle
\begin{abstract}
Effective ultrasound tissue characterization is usually hindered by complex tissue structures. The interlacing of speckle patterns complicates the correct estimation of backscatter distribution parameters. Nakagami parametric imaging based on localized shape parameter mapping can model different backscattering conditions. However, performance of the constructed Nakagami image depends on the sensitivity of the estimation method to the backscattered statistics and scale of analysis. Using a fixed focal region of interest in estimating the Nakagami parametric image would increase estimation variance. In this work, localized Nakagami parameters are estimated adaptively by means of maximum likelihood estimation on a multiscale basis. The varying size kernel integrates the goodness-of-fit of the backscattering distribution parameters at multiple scales for more stable parameter estimation. Results show improved quantitative visualization of changes in tissue specular reflections, suggesting a potential approach for improving tumor localization in low contrast ultrasound images. 
\end{abstract}
\begin{keywords}
Nakagami imaging, tumor detection, maximum likelihood estimation, liver tumor, RF envelope
\end{keywords}
\section{Introduction}
\label{sec:intro}

Ultrasound parametric imaging is gaining increased interest as an effective way for quantitative tumor characterization. Changes in properties of soft tissue texture, e.g. liver parenchyma, can be reflected in the radio-frequency (RF) backscattered statistics as different Rayleigh distributions \cite{mau13}. However parametric estimation is not a straightforward process and is generally faced with increased estimation variance, especially for complex speckle patterns. This may obscure abnormal tissue structures, e.g. tumors and fibrosis, which are deemed important for early diagnosis \cite{hyt09}.

The analytical simplicity of the bi-parametric Nakagami distribution model, along with its goodness-of-fit with the envelope histogram of the ultrasound-backscattered signal \cite{snk00}, can be attractive for tissue characterization \cite{mau13,alk15,alk16}. The shape of the Nakagami distribution is specified by the $\mu$ parameter corresponding to the local concentration of scatterers, and the amount of spread (i.e. the local backscattered energy) is represented by the scale parameter $\omega$. Different conditions of the RF envelope statistics can be achieved by varying the $\mu$ parameter. Values of $\mu$ between 0 and 1 yield pre-Rayleigh and Rayleigh distributions. The Rayleigh distribution case $\left(\mu = 1\right)$ resembles of having a large number of randomly distributed scatterers, and in the case of high degree of variance the distribution conforms to pre-Rayleigh $\left(\mu < 1\right)$. For a mixture of random and periodically located scatterers, the RF envelope statistics becomes a post-Rayleigh distribution $\left(\mu > 1\right)$. The map of local $\mu$ parameter values -- that correspond to tissue properties -- is normally considered in constructing the Nakagami parametric image. The estimated Nakagami parameters as a function of the backscattered envelope statistics has shown to be a reliable tool for quantitative visualization of tissue structure changes \cite{alk15,alk16,tsi15}.\\
\indent Previous work has improved local window-based $\mu$ parameter estimation to generate the Nakagami parameter map from envelopes of raw ultrasound signals \cite{lru11,ho12,tsi14}. As using a gamma kernel density estimation to achieve a smooth estimation of the distribution from small fixed-size windows \cite{lru11}, or by using a number of windows having a size 3 times the pulse length of the ultrasound \cite{ho12}, or by summing and averaging multiple Nakagami parametric images generated using different sliding square window sizes (7-10 times the transducer pulse length) \cite{tsi14}. However challenges persist with fixed-size window approaches. A focal region of interest (i.e. using small windows) enables enhanced resolution of the Nakagami parametric image, but large tissue structures require a large spatial scale to achieve stable parameter estimation. Therefore parameter smoothing might not suffice when prominent parts of the examined tissue structure is truncated or located outside the window borders. On the other hand, using large window sizes for summation and averaging may affect the results when compounded with windows of smaller-sizes, and hence affecting the reliability of the constructed parametric image resolution.\\
%\indent For example, given an arrangement of pixels in a focal image region $I\left(x,y\right)$ defined over a discrete grid indexed by $l \in \mathbb{Z}^2$, Fig.1 shows a localized kernel $v\left(I\left(x+s,y+t\right)\right)$ applied to $I\left(x,y\right)$ at position $s,t \in \mathbb{R}^2$. Each of the processed outputs covered by $v$, depend on a local region $m \times n$ which is a subset of $x,y \in M \times N$. This fixed focal region approach could be not sufficient to cope with the various texture structures within $I\left(x,y\right)$, as each region would require a varying size of v depending on the texture structure size and density. Therefore the aggregation of the outputs of $v \forall x,y \in M \times N$ forming the parametric image, would not accurately represent tissue characteristics. This situation is commonly faced with the heterogeneous liver tissue texture -- the point of emphasis in this work.
\indent In this work an alternative approach of employing a multiscale kernel-based technique to model the backscattering distribution statistics is proposed. The focal region of interest should be large enough to have sufficient tissue variation, while being also as small to avoid inclusion of irrelevant textures from the nearby regions. The backscattered envelope from tissue was estimated voxel-by-voxel via Nakagami distribution and subsequently used to generate optimized local parametric images for improved liver tumor detection. The assumption is based on that tumor regions tend to have different backscattered distribution than normal tissue \cite{trl00}, and a localized approach based on a varying-size kernel can assist in better identifying the tumor speckle patterns.%; especially in its early stages when the differences are subtle.

%\begin{figure}[t]

%\begin{minipage}[b]{1\linewidth}
%  \centering
%  \centerline{\includegraphics[width=8.5cm]{Figures/Fig1}}
%%  \vspace{2.0cm}
%  %\centerline{}\medskip
%\end{minipage}
%%
%\caption{[Right] An illustrative example of the need of a varying size ${m}\times{n}$ kernel $v$ to better localize the various  structures in image texture, and [Left] different kernel sizes for corresponding image texture.}
%\label{fig:res1}
%
%\end{figure}

\section{Methodology}
\label{sec:method}

The speckle pattern is dependent on the ultrasound wavelength and underlying tissue structure. As the former factor is fixed in this case and the latter varies across the ultrasound image, a multiscale approach that can localize the different tissue structures would best suit the modeling of the backscattered envelope. A non-linear kernel is applied adaptively to characterize the tissue speckle patterns. The estimation of the best parametric Nakagami maps from varying size kernels is described as follows:

\subsection{Multiscale kernel localization}
Let $V$ be an order set of constructed envelope detected RF images $I_{i}\left(x, y\right)$, where $i$ is a certain slice in the acquired volume, and $P_{\mu,\omega}(x,y)$ are the corresponding $\mu$ and $\omega$ parametric images. The RF images are calculated from the envelope of the ultrasound-backscattered signal just before performing any intensity mapping and post processing filtering. This representation preserves the ultrasound data unaltered while providing better quantitative analysis, i.e. without the risk of losing information due to RF signal shaping. Then a set of varying size kernels $\mathscr{K}$ can be defined for each $I_{i}\left(x, y\right)$, where $\mathscr{K} =\left\{v_{1},v_{2},\ldots,v_{k}\right\}, v_{j} \in V$ and $k = 1/8$ of the size of $I_{i}\left(x, y\right)$. Different focal regions are investigated in a multiscale manner as in (1) by varying the size of two non-negative integer variables $a$ and $b$, which are used to center each localized kernel $v_{j}(s, t)$ with a different size of $m \times n$ on each voxel $l$ in $I_{i}\left(x, y\right)$ of size $M \times N$.
\begin{equation}
P_{\mu,\omega}(x,y)= \sum_{s=-a}^{a}\sum_{t=-b}^{b}v_{j}\left(s,t\right)I_{i}\left(x+s,y+t\right)\left(\frac{k}{j}\right)^2
\end{equation}
where $a=\ceil{\frac{m+2}{2}}$, $b=\ceil{\frac{n+2}{2}}$, and $m$, $n = 1,2,\dots,k$.

\subsection{Modeling backscattered statistics}

The Nakagami distribution $N(x)$ is a gamma related distribution which is known for its analytical simplicity \cite{alk16}, and has been proposed as a general model for ultrasonic backscattering under different scattering conditions and scatterer densities \cite{snk00}. This distribution has the density function

\begin{equation}
N(x|\mu,\omega)=2\left(\frac{\mu}{\omega}\right)^{\mu}\frac{1}{\Gamma\left(\mu\right)} x^{\left(2\mu-1\right)} e^{-\frac{\mu}{\omega}x^{2}} ,  \quad \forall x \in \mathbb{R} \geq 0
\end{equation}

\noindent where $x$ is the envelope of the RF signal and $\Gamma\left(\cdot\right)$ is the gamma function. If $x$ has a Nakagami distribution $N(x)$ with parameters $\mu$ and $\omega$, then $x^2$ has a gamma distribution $\Gamma$ with shape $\mu$ and scale (energy) parameter $\omega$/$\mu$. Although there are other distributions exist in the literature for modeling ultrasonic backscattering, the Nakagami probabilistic distribution was chosen for its simplicity and ability to characterize different scattering conditions ranging from pre- to post-Rayleigh \cite{tsi15}.

Each voxel $l$ in $I_i$ is adaptively transformed via $\mathscr{K}$ at different scales to its corresponding parametric Nakagami parameters by means of maximum likelihood estimation (MLE) forming a set of parametric vectors for each voxel $l$. The MLE $\hat{\theta} \left(v\right)$ for a density function $f\left(v^l_{1},\ldots,v^l_{k} | \theta\right)$ when $\theta$ is a vector of parameters for the Nakagami distribution family $\Theta$, estimates the most probable parameters $\hat{\theta}\left(v\right) = arg max_\theta \: D\left(\theta|v^l_{1},\ldots,v^l_{k}\right)$, where $D\left(\theta |v\right) = f\left(v|\theta\right), \! \theta \in \Theta$ is the score function. Finally the goodness-of-fit is estimated via root mean square error for the calculated Nakagami parameters $\theta_m$ at different scales, giving the localized parametric Nakagami images $P_{\mu,\omega}$ as summarized in Algorithm 1. The shape parametric image $P_{\mu}$ is used for subsequent tissue characterization.

%This gives more reasoning in analyzing the different ultrasound tissue regions, as different regions tend to exhibit a mixture of distributions. Characterizing the speckle patterns via different localized focal regions of interest could assist in better representing the backscattered signal, which adapts according to the varying nature of tissue texture.

\begin{algorithm} [!ht]
  \caption{Multiscale kernel localization}
  \begin{algorithmic}
    \REQUIRE Set of ultrasound backscattered envelope images $I_i = \left\{\left(x_1,y_1\ldots,x_{j},y_{j} \right)\right\}$
    \ENSURE Localized ultrasound Nakagami shape and scale parametric images $P_{\mu}, P_{\omega}$
    \FORALL {voxels $l$ in $I_i$} % Outer loop
    \FORALL {localized kernels $\nu_{1}^{l} \rightarrow \nu_{k}^{l}$}
		    %\bigskip% STEP 1
				\STATE \COMMENT {\textbf{Step1}}
				//\underline{ Fit with a Nakagami distribution}
				$N(x|\mu,\omega)=2\left(\frac{\mu}{\omega}\right)^{\mu}\frac{1}{\Gamma\left(\mu\right)} x^{\left(2\mu-1\right)}e^{-\frac{\mu}{\omega}x^{2}}$
			%\bigskip% STEP 2
				\STATE \COMMENT {\textbf{Step2}} //\underline{ Calculate Nakagami shape $\mu$ and scale $\omega$}
				\underline{parameters using maximum likelihood estimation as:}
				\STATEx $\hat{\theta}\left(\nu\right) = arg max_\theta \: D\left(\theta/\nu^l_{1},\ldots,\nu^l_{k}\right)$
				\STATEx where $\theta$ is a vector of parameters for the Nakagami distribution family $f\left(\nu^l_{1},\ldots,\nu^l_{k}/\theta\right)$
	\ENDFOR
				%\bigskip% STEP 3
				\STATE \COMMENT {\textbf{Step3}}
				//\underline{ Estimate goodness-of-fit of the determined}
				\underline{Nakagami parameters $\theta_{m}$ with the average RF signal $\theta_{\alpha}$ }
				\underline{within set of localized kernels $\mathscr{K}$}
				
				$\left(P_{\mu_{1},\omega_{1}},\dots, P_{\mu_{j},\omega_{j}}\right) =  \argmin\left\{\sqrt\frac{\sum\limits_{s=2}^n \left(\theta_{m} - \theta_{\alpha}\right)^2} {n}\right\}$
	\ENDFOR
	\RETURN{ $P_{\mu}, P_{\omega}$} 
	\end{algorithmic}
\end{algorithm}

\begin{figure}[t]

\begin{minipage}[b]{1.0\linewidth}
  \centering
  \centerline{\includegraphics[width=5.0cm]{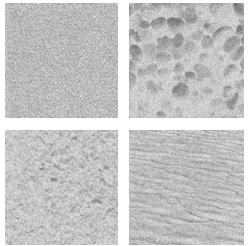}}
%  \vspace{2.0cm}
  %\centerline{(a) Result 1}\medskip
\end{minipage}
\caption{Samples of simulated ultrasound speckle images representing: [left-right] fine texture (dense scatterers), coarse texture (sparse scatterers), heterogeneous texture (random scatterers), and homogeneous texture (periodic scatterers), referring to, respectively, phantoms D57, D30, D5, D37 in Table~\ref{table:table1}.}
\label{fig:fig1}
\end{figure}

\begin{table}
\centering
\caption{Mean absolute difference comparison 
of estimated Nakagami shape parametric 
images against ground-truth. \newline}
\label{table:table1}
\begin{tabular}{llll}
 \hline \hline
                             &  \multicolumn{3}{c}{\cellcolor[HTML]{B3CAE3}{\color[HTML]{333333} Nakagami shape}}     \\ 
\multirow{-2}{*}{}          & \multicolumn{3}{c}{\cellcolor[HTML]{B3CAE3}{\color[HTML]{333333} estimation methods}} \\  \hline
\rowcolor[HTML]{D9E4F0} 
Phantom                      & GKF                   & WMC                           & MKL                           \\ \hline
\cellcolor[HTML]{D9E4F0}D5   & 0.29                  & 0.10                          & \textbf{0.04}                 \\ \hline
\cellcolor[HTML]{D9E4F0}D11  & 0.18                  & 0.09                          & \textbf{0.04}                 \\ \hline
\cellcolor[HTML]{D9E4F0}D13  & 0.40                  & 0.17                          & \textbf{0.01}                 \\ \hline
\cellcolor[HTML]{D9E4F0}D30  & 0.51                  & 0.33                          & \textbf{0.07}                 \\ \hline
\cellcolor[HTML]{D9E4F0}D37  & 0.29                  & 0.09                          & \textbf{0.05}                 \\ \hline
\cellcolor[HTML]{D9E4F0}D57  & 0.30                  & \textbf{0.03}                 & 0.07                          \\ \hline
\cellcolor[HTML]{D9E4F0}D71  & 0.16                  & 0.09                          & \textbf{0.02}                 \\ \hline
\cellcolor[HTML]{D9E4F0}D88  & 0.64                  & 0.46                          & \textbf{0.12}                 \\ \hline
\cellcolor[HTML]{D9E4F0}D91  & 0.55                  & 0.47                          & \textbf{0.19}                 \\ \hline
\cellcolor[HTML]{D9E4F0}D99  & 0.49                  & 0.28                          & \textbf{0.04}                 \\ \hline
\cellcolor[HTML]{D9E4F0}D101 & 0.62                  & 0.13                          & \textbf{0.11}    \\ \hline \hline            
\end{tabular}
\end{table}

\section{RESULTS}
\label{sec:results}

\subsection{Simulated ultrasound speckle images}
Simulation experiments were performed on 11 different ultrasound speckle images generated from corresponding texture images adopted from the Brodatz texture album \cite{bdz96}. The ultrasound speckle images were synthesized with given textures as the initial point scatterer image, giving clinical echo \textit{alike} images that resemble tissue scatterers in appearance. Various specular reflection conditions of tissue texture boundaries are synthesized ranging from fine to coarse (i.e. high density to low density scatterers) and from heterogeneous to homogeneous (i.e. random to periodic scatterers alignment), c.f. Fig.~\ref{fig:fig1}.

The window-based $\mu$ parameter estimation methods: gamma kernel function (GKF) \cite{lru11}, windows-modulated compounding (WMC) \cite{tsi14} and the proposed multiscale kernel localization (MKL) methods where applied to the simulated ultrasound speckle images, and performance quantitatively compared with the $\mu$ parameters estimated from the original synthetic texture  images (ground-truth), as shown in Table~\ref{table:table1}. Results show that the MKL method gives more stable $\mu$ parameter estimation in nearly all cases.

\subsection{Liver tumor detection}
\begin{figure}[t]

\begin{minipage}[b]{0.48\linewidth}
  \centering
  \centerline{\includegraphics[width=3.25cm]{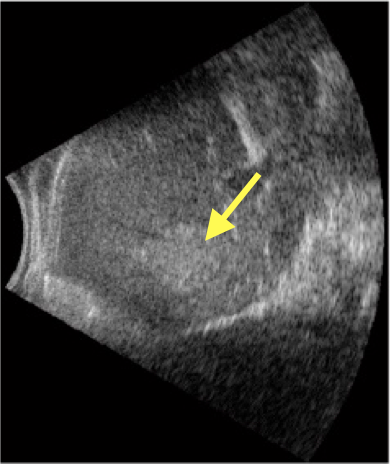}}
%  \vspace{2.0cm}
  \centerline{(a)}\medskip
\end{minipage}
\hfill
\begin{minipage}[b]{0.48\linewidth}
  \centering
  \centerline{\includegraphics[width=4.0cm]{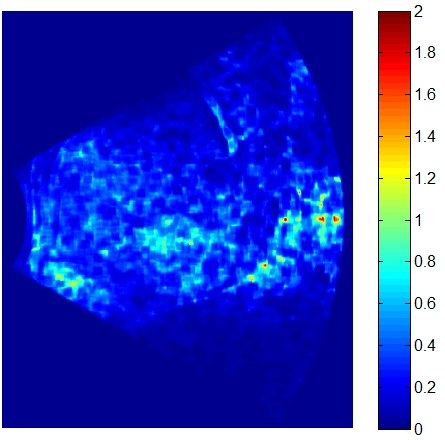}}
%  \vspace{1.5cm}
  \centerline{(b)}\medskip
\end{minipage}
\begin{minipage}[b]{0.48\linewidth}
  \centering
  \centerline{\includegraphics[width=4.0cm]{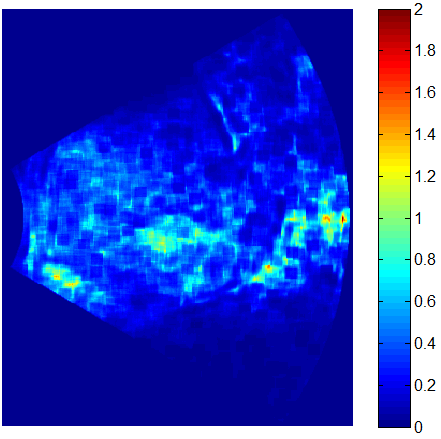}}
%  \vspace{1.5cm}
  \centerline{(c)}\medskip
\end{minipage}
\hfill
\begin{minipage}[b]{0.48\linewidth}
  \centering
  \centerline{\includegraphics[width=4.0cm]{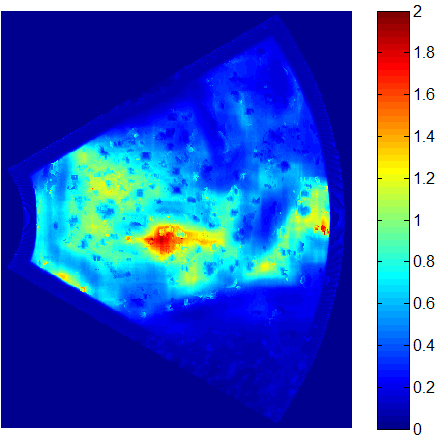}}
%  \vspace{1.5cm}
  \centerline{(d)}\medskip
\end{minipage}
\caption{(a) Clinical ultrasound B-mode image showing a liver tumor (indicated by a yellow arrow), and corresponding Nakagami shape parametric image using (b) WMC, (c) GKF, and (d) MKL methods.}
\label{fig:fig2}
\end{figure}

\begin{figure}

\begin{minipage}[b]{1.0\linewidth}
  \centering
  \centerline{\includegraphics[scale=0.75]{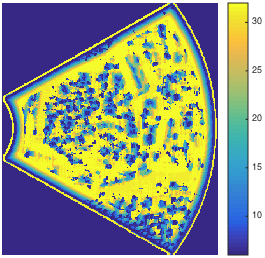}}
%  \vspace{2.0cm} [width=5.0cm]
  %\centerline{(a) Result 1}\medskip
\end{minipage}
\caption{Localized adaptive kernel sizes for Fig.~\ref{fig:fig2}(d).}
\label{fig:fig3}
\end{figure}

In order to quantitatively evaluate the robustness of the 3 different Nakagami parametric image estimation methods, they were applied to real ultrasound liver tumor images obtained using a diagnostic ultrasound system (z.one, Zonare Medical Systems, Mountain View, CA, USA) with a 4 MHz curvilinear transducer and 11 MHz sampling. The whole RF ultrasound image (without log-compression and filtering) was used in generating the Nakagami parametric images, so the sensitivity of methods to various tissue scatterers can be investigated. Fig.~\ref{fig:fig2} shows an ultrasound liver tumor image and corresponding Nakagami parametric images via the 3 different methods. Tumor tissue specular reflections tend to appear more prominent from the background tissue using the MKL approach as compared to GFK and MWC window-based $\mu$ parameter estimation methods. The different kernel sizes used in generating the Nakagami parametric image using the MKL method is shown in Fig.~\ref{fig:fig3}.

%\vfill
%\pagebreak

\section{DISCUSSION}
\label{sec:discussion}
Tumor texture tends to be more heterogeneous as compared to normal tissue \cite{alk09b}. This property has been reported to be useful in tumor grading \cite{alk09b,alk09a,alk10,alk15b} and assessing aggressiveness \cite{alk08}. However tumor spatial and contrast resolution in ultrasound images is low as compared to other modalities. Modeling the RF backscattered envelope from liver tissue requires an adaptive method that can effectively investigate tissue heterogeneity while reliably estimate the distribution parameters. Different spatial variations exist in speckle patterns across the ultrasound image due to the Rayleigh scattering behavior \cite{mau13}, and many tissue structures are prone to low spatial contrast and displacement during successive image acquisition. This makes the use of constant focal regions in estimating the backscattering distribution parameters very limiting. Such approach may result in missing parts of the analyzed speckle pattern if the focal region was too small, or possibly interlacing of irrelevant patterns from surrounding regions if the focal region was too large. Thereby subtle tissue structure (e.g. tumor regions in its early stages) could be obscured due to the presence of mixture of patterns.\\% and hence rendered harder to detect.\\
\indent Different window sizes have diverse effects on the formation of the Nakagami parametric image. The experiments on simulated and real ultrasound images demonstrated the need for an adaptive approach that can enhance image resolution without degrading smoothness, i.e. having stable parameter estimation. Stable performance was achieved using the MKL method when applied to diverse speckle patterns simulating different soft tissue conditions in clinical practice. An exceptional case of D57 in Table~\ref{table:table1} -- which had a fine tissue structure -- did not give the best stable $\mu$ estimation. The uniform speckle pattern appearance across the D57 image texture would reduce the sensitivity to texture variations of the adaptive approach employed by MKL. Thus a variant spatial resolution throughout the entire imaging field of view would not be best for this particular case \cite{xue15}. However in clinical practice, liver tissue characterization involves analyzing the whole ultrasound image before the tumor is localized (cf. Fig.~\ref{fig:fig2}(a)), which means encountering regions with different tissue characteristics; thus a non-varying window size may reduced the reliability of Nakagami imaging.\\
\indent The ability to rapidly and accurately identify tumor location in ultrasound images is limited due to inherent low contrast. Therefore ultrasound parametric imaging is normally applied to analyze the RF envelope statistics to give an indication of the properties of tissue scatterers. Fig.~\ref{fig:fig2}(d) shows visual improvement in the contrast between the specular reflections of tissue boundaries (c.f. Fig.~\ref{fig:fig2}(b) and Fig.~\ref{fig:fig2}(c)), with a stronger parametric response in the localized tumor region from surrounding tissue. This could be attributed to the adaptive approach of the MKL method that integrates the goodness-of-fit of the backscattering distribution parameters at multiple scales before parameter estimation. Examining how the focal regions vary in size as shown in Fig.~\ref{fig:fig3}, the MKL allows for the aggregation of sufficient voxels that would better represent the envelope statistics in order to highlight differences in tissue properties. Such localized multiscale neighborhood around each voxel contributes for the best resolution and improved smoothness.\\ 
\indent Finally, a number of challenges may arise with the employment of Nakagami imaging in tumor segmentation, such as the presence of blood vessels, ducts and other connective tissues. Although these small areas might give signs on liver inflammation, they would rather degrade the local image resolution and hence affect the smoothness of the parameter estimation. Also tumor spatial contrast varies according to depth and level of speckle artifacts. Such challenges would serve as future work for improving accurate segmentation of tumor boundaries in Nakagami parametric images.

\section{CONCLUSION}
\label{sec:conclusion}
Nakagami parametric imaging based on localized shape parameter maps can model different backscattering conditions. Results show more stable estimation of the backscattering distribution parameters within a varying size kernel by means of MLE. Moreover, improved highlighting of tumor tissue specular reflections in ultrasound images was achieved. The proposed technique could serve as a decision support tool to model the statistical distribution of ultrasound backscatter signals for improved detection of liver tumors.

%\begin{figure}[t]
%
%\begin{minipage}[b]{0.48\linewidth}
%  \centering
%  \centerline{\includegraphics[width=4.5cm]{Figures/Fig4-a}}
%%  \vspace{2.0cm}
%  \centerline{(a)}\medskip
%\end{minipage}
%\hfill
%\begin{minipage}[b]{0.48\linewidth}
%  \centering
%  \centerline{\includegraphics[width=4.5cm]{Figures/Fig4-b}}
%%  \vspace{1.5cm}
%  \centerline{(b)}\medskip
%\end{minipage}
%%
%\begin{minipage}[b]{1.0\linewidth}
%  \centering
%  \centerline{\includegraphics[width=3.9cm,height=3.9cm]{Figures%/Fig4-c}}
%%  \vspace{1.5cm}
%  \centerline{(c)}\medskip
%\end{minipage}
%%
%\caption{(a) Different localized kernel regions within the tumor %tissue indicated in Fig. 2(a), and corresponding Nakagami scale %and shape parametric images using MKL method is shown in (b) and %(c), respectively.}
%\label{fig:fig4}
%%
%\end{figure}

% BiBTeX files (here: refs). The IEEEbib.bst bibliography
\bibliographystyle{IEEEbib}
\bibliography{main_ICIP}

\end{document}